# Radiation Tolerance of Low-Noise Photoreceivers for the LISA Space Mission

P. Colcombet[(1)(2)], N. Dinu-Jaeger[(1)], C. Inguimbert[(2)], T. Nuns[(2)], S. Bruhier[(1)], N. Christensen [(1)] P. Hofverberg[(3)], N. van Bakel[(4)], M. van Beuzekom[(4)], T. Mistry[(4)], G. Visser[(4)], D. Pascucci[(4)], K. Izumi [(5)], K. Komori[(5)], G. Heinzel[(6)], G. Fernández Barranco[(6)], J.J.M. in t Zand[(7)], P. Laubert[(7)], M. Frericks[(7)]

(1) Université Côte d'Azur, Observatoire de la Côte d'Azur, CNRS, ARTEMIS, 06304 Nice, France
(2) ONERA The French Aerospace Lab -DPHY, 31055 Toulouse, France
(3) Institut Méditerranéen De Protonthérapie – Centre Antoine Lacassagne Nice, France
(4) Nikhef Dutch National Institute for Subatomic Physics, Amsterdam, The Netherlands
(5) Japan Aerospace Exploration Agency (JAXA) Tokyo, Japan
(6) Albert Einstein Institute (AEI) Hannover, Germany
(7) SRON Netherlands Institute for Space Research Leiden, Netherlands

*Abstract*— This study investigates the effects of space environmental radiation on the performance of InGaAs Quadrant Photodiodes (QPDs) and assesses their suitability for the Laser Interferometer Space Antenna (LISA) mission. QPDs of 1.0, 1.5 and 2.0 mm have been irradiated with 20 and 60 MeV protons, 0.5 and 1 MeV electrons, and Co$^{60}$ gamma. An exposure corresponding to a displacement damage equivalent fluence of $1.0 \times 10^{12}$ p/cm² for 20 and 60 MeV protons and a total ionizing dose of 237 krad were applied, surpassing the anticipated radiation levels for the LISA mission by a factor of approximately five. Experiments were conducted to measure changes in QPD dark current, capacitance, and responsivity. The QPDs are integrated with a low-noise DC-coupled transimpedance amplifier to form the Photoreceiver (QPR). QPR noise and performance in an interferometric system like LISA were also measured. Although radiation impacted their dark current and responsivity, almost all QPDs met LISA's validation criteria and did not demonstrate any critical failure. These findings prove that the tested QPDs are promising candidates for LISA and other space-based missions.

## I. INTRODUCTION

The Laser Interferometer Space Antenna (LISA) mission planned for 2037 and led by the European Space Agency (ESA), will be the first space-based low-frequency gravitational-wave (GW) detector, operating in the frequency range of $10^{-4}$ to 1 Hz. This low-frequency band, inaccessible to ground-based detectors, has the potential to open new frontiers in the study of the universe [1]–[4].

LISA comprises a cluster of three spacecraft, arranged in an equilateral triangle with 2.5 million km side, situated in the Earth-like heliocentric orbit at 1 astronomical unit from the Sun, and located at an angle of 20° behind the Earth. The detection of GWs is accomplished by measuring the intra-spacecraft distance variations between free-fall test masses positioned in separate spacecraft, using high-precision laser interferometers operating at a wavelength of 1064 nm [5].

Identical photoreceivers, known as Quadrant Photoreceivers (QPRs), are critical components in LISA's design, responsible for detecting interference between optical signals of disparate intensities, ~700 $pW$ and 1 $mW$. A QPR is composed of an InGaAs quadrant photodiode (QPD) connected to a low-noise DC-coupled transimpedance amplifier (TIA) assembled in a mechanical housing. The QPDs are custom devices developed by the Netherlands (NL) and Japan (JP). The current version of the TIA is designed by Germany (DE) [6]. To validate LISA specifications, the QPD must demonstrate a responsivity > 0.7 A/W, a dark current < 1 μA and a 1.5 mm diameter active area to increase the sensitivity of the detector and improve the alignment. The equivalent input current noise of the QPR must be maintained below 2 pA/√Hz over the LISA heterodyne frequency range from 5 to 30 MHz.

During an extended 12.5-year mission, LISA is exposed to various forms of radiation such as solar wind, solar flares, and cosmic radiation. Energetic particles from these sources penetrate the spacecraft walls and interact with matter either through the atomic electron cloud or nuclei. The former type can ionize the atoms and create electron-hole pairs, leading to a cumulative effect known as Total Ionizing Dose (TID) quantified by the Linear Energy Transfer (LET). The latter interaction can eject nuclei and create vacancies and interstitials in the semiconductor lattice determining atomic displacement damage [7], [8]. This process is quantified by the Non-Ionizing Energy Loss (NIEL) and Displacement Damage Dose (DDD) [9]. Both NIEL and LET are defined for a given energy particle in a given material.

In the LISA orbit, solar flare events constitute the primary radiation challenge. For a 12.5-year mission with a 3 mm aluminium shield, ESA's worst-case scenario requirements for the optical bench components, a TID of 40 krad (InGaAs) and a Displacement Damage Equivalent Fluence (DDEF) of $1.01 \times 10^{+11}$ p/cm² for 10 MeV protons, corresponding to a DDD of $6.62 \times 10^{+8}$ MeV/g (InGaAs). Using the mission's cumulative fluence from the NASA GSFC ESP model [10], and InGaAs NIEL data from NEMO [11], the software OMERE [12] can be employed to re-evaluate the 10 MeV proton equivalent fluence, yielding results consistent with the original DDEF. Additionally, OMERE provides the DDEF for 20 and 60 MeV energies, which are relevant for proton irradiation, as depicted in Figure 1.

These fluences of particles are significant enough to induce defects and affect the QPD and QPR performances. Therefore, this study aims to assess the tolerance of QPD performance with respect to dark current, capacitance, and responsivity under proton, electron, and gamma irradiation at LISA fluences and beyond. Additionally, the investigation will include QPR noise, and optical performance regarding amplitude and phase in a system, reproducing LISA interferometers. This research will contribute to a comprehensive understanding of the impact of radiation on QPD performance and the overall interferometric system.

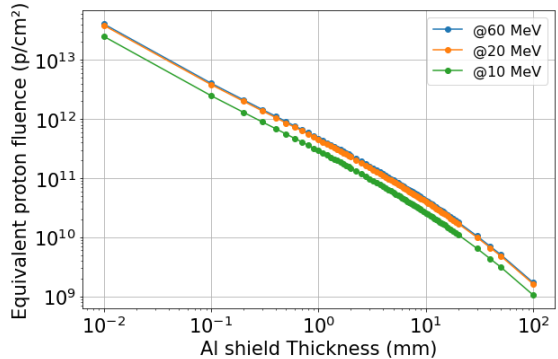

*Figure 1 OMERE proton equivalent fluence for LISA's 12.5-year lifetime, by Al shield thickness and energy*

## II. IRRADIATION DESCRIPTION

### A. Irradiation conditions and facilities

The QPDs were exposed to radiation using protons, electrons, and gamma rays as detailed in Table I and Table II. The highest TID reached during these exposures was 237 krad. Additionally, the cumulated fluence for 20 MeV and 60 MeV protons was $1.0 \times 10^{12}$ p/cm². Both surpass the stipulated LISA requirements for TID and DDD by a factor of approximately five.

Proton irradiation was carried out using the MEDICYC R&D line at the Centre Antoine Lacassagne (CAL) (Nice, France) [13]. The facility hosts a 65 MeV isochronous cyclotron which delivers a 10 cm diameter beam with homogeneity < ±3%. Electron irradiation was performed at ONERA (Toulouse, France) in the MIRAGE facility. The beam is supplied by a 2 MeV electron Van de Graaff accelerator. Flux is measured with a Faraday cup. The overall homogeneity and accuracy are ±10%. Samples were exposed under a vacuum of $10^{-6}$ to $10^{-7}$ mbar. Gamma irradiations were conducted also at ONERA, using the $Co^{60}$ MEGA facility. The dosimetry is performed with ionisation chambers with the same overall accuracy of ±10%.

*Table I TID & DDD applied on the QPD for all irradiations.*

| Particle type | Energy (MeV) | Fluence (p/cm²) | TID (krad) | DDD (MeV/g) |
|---|---|---|---|---|
| Proton (CAL) | 20 | 1x10+12 | 237 | 4.9x10+9 |
|  | 60 | 1x10+12 | 104 | 3.6x10+9 |
| Electron (ONERA) | 0.5 | 5x10+12 | 105 | 3.36x10+7 |
|  | 1 | 5x10+12 | 100 | 9.71x10+7 |
| Gamma (ONERA) | 1.25 |  | 237 | ~7.6x10+7 |

*Table II Irradiated QPD organization.*

| Type | Energy (MeV) | Total Number of QPDs | QPD Origin | QPD Size |
|---|---|---|---|---|
| Protons (CAL) | 20 | 3 | JP | 1, 1.5 & 2 mm |
|  |  | 1 | NL | 1.5 mm |
|  | 60 | 3 | JP | 1, 1.5 & 2 mm |
|  |  | 1 | NL | 1.5 mm |
| Electrons (ONERA) | 0.5 | 2 | JP | 1 & 2 mm |
|  | 1 | 3 | JP | 1, 1.5 & 2 mm |
| Gamma (ONERA) | 1.25 | 1 | JP | 1.5 mm |
|  |  | 1 | NL | 1.5 mm |

*Table III Irradiation fluence step for Proton and Electron radiation and cumulative Dose for Gamma radiation*

| Proton Cumulative Fluence (p/cm²) | | Electron Cumulative Fluence (e/cm²) | Gamma Cumulative Dose (krad) |
|---|---|---|---|
| 60 MeV | 20 MeV | 0.5 & 1 MeV | 1.25 MeV |
| 2x10+9 | 2x10+9 | 0,5 10+12 | 1 |
| 4x10+9 | 4x10+9 | 1x 10+12 | 2 |
| 8.4x10+9 | 8.4x10+9 | 2x10+12 | 22 |
| 4.2x10+10 | 4.2x10+10 | 3x10+12 | 36 |
| 2.1x10+11 | 1x10+12 | 4x10+12 | 108 |
| 1x10+12 |  | 5x10+12 | 150 |
|  |  |  | 237 |

### B. Description of Pre-, Post-, and In-Situ tests

Prior to and after irradiation, measurements were conducted for dark current, capacitance, and responsivity across all QPDs as well as the QPR input current noise. Additionally, the amplitude and phase of the QPR in response to a LISA-type signal further elaborated in Section C, were assessed on four specific QPDs: two NL QPDs (one as a reference and the other irradiated) and two JP QPDs (one as a reference and the other irradiated), each having a 1.5 mm diameter. The fluence steps during the irradiation campaigns are detailed in Table III. Between each step, only variations in capacitance at 25°C, dark current at 20, 35 and 50°C, as well as noise were recorded.

The dark current and capacitance of the QPD are measured using a KEITHLEY 4200 SCS semiconductor characterization system. The QPD is placed in a temperature-controlled copper support, which is housed in a metallic enclosure to ensure darkness. Dark current is simultaneously measured on all quadrants with the same reverse bias voltage applied to each quadrant at different temperatures between 20°C to 50°C. The capacitance of each quadrant is individually measured at 25°C and with a voltage sweep like for the dark current measurements and a frequency sweep of 1 to 10 MHz. The internal offset capacitance and current from the setup (cable, instruments, etc.) are evaluated and subtracted from the results presented in Figure 4 and Figure 5.

Responsivity is defined as the photocurrent-to-optical power ratio of a photodiode. The setup used is illustrated in Figure 2. A 1064 nm continuous laser beam, produced by a fibered laser source, is focused onto a QPD quadrant using a collimator. The laser beam's shape and size are prior measurements, calibrated using a Thorlabs beam profiler. The beam's optical power is tracked using a NIST InGaAs photodiode, and the QPD's photocurrent is measured using a Keithley 2600B source meter.

The input current noise and the transimpedance amplifier

(TIA) gain of each channel in the QPR are measured using the "white light method", which is described in detail in reference [14] (p. 79) and [15]. The latter reference reports the same experimental setup, with the exception that we additionally measure the noise floor of the measurement chain, in addition to the spectral noise in dark and light conditions using a Rohde and Schwarz FSP13 spectrum analyser.

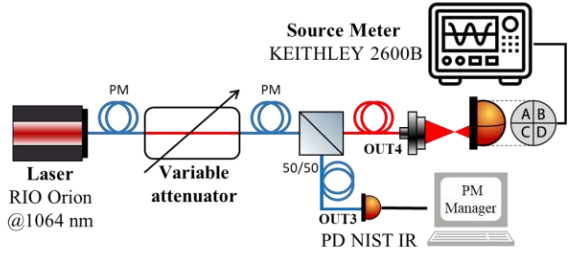

*Figure 2 Experimental responsivity measurement setup*

### C. Experimental setup for QPR characterization under equivalent interferometric optical signal

The LISA mission employs three interferometers: the long arm, the test mass, and the reference interferometer. Each LISA interferometric optical signal $S(t)^{HET}$ can be expressed as equation (1), which describes the interaction of two coherent laser beams with optical powers $P_1$ and $P_2$ and efficiency $\eta_e$ of the heterodyne interferometer. This interaction results in a continuous component $P_{DC} = (P_1 + P_2)/2$ and an RF one, $P_{AC} = \sqrt{\eta_e P_1 P_2} \sin(\varphi_r(t))$ where $\varphi_r(t)$ is the phase difference between the two laser beam.

$$S(t)^{HET} = \frac{P_1 + P_2}{2} + \sqrt{\eta_e P_1 P_2} \sin(\varphi_r(t)) \quad (1)$$

To accurately reproduce each LISA interferometer, our experimental setup (see Figure 3) employs an intensity modulator controlled with a Zurich lock-in amplifier [16]. The output signal is monitored using two photodiodes, a NIST photodiode for the DC optical signal and a Thorlabs calibrated photodiode with a 1.2 GHz bandwidth for the AC signal. The laser beam is focused on a single QPD segment. The amplitude and phase response of the illuminated QPR channel are observed relative to an optical power sweep from 0 to 110 µA, and a frequency sweep from 0 to 30 MHz representative of LISA optical power and bandwidth.

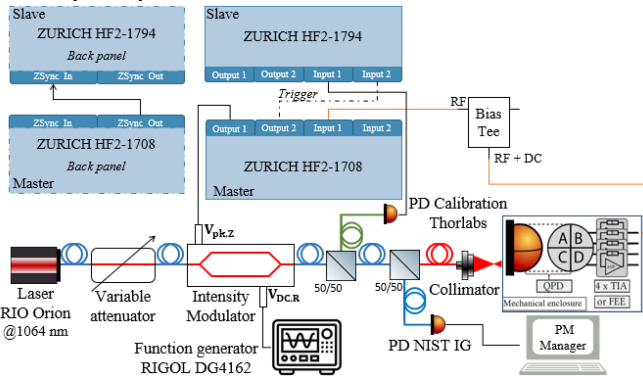

*Figure 3 Experimental setup for measuring QPR phase and amplitude response to AC signals.*

## III. DATA RESULT AND ANALYSIS

### A. Pre-irradiation results

The QPDs pre-irradiation meet LISA standards with dark currents between 50 pA and 1 nA at full depletion voltage and at 20°C. QPDs with 1.5 mm and 1 mm active areas maintain noise levels within LISA's 2 pA/√Hz limit across 5-30 MHz (see Figure 4). In addition, a clear relationship emerges between QPD's active area, capacitance, and QPR noise, as illustrated in Figure 5. While variations in capacitance between channels in the NL QPDs were detected in our measurements, the NL team reported no deviation, suggesting potential nuances in the experimental setups. In Figure 9, pre-irradiation responsivity results display values up to 0.83 A/W exceeding LISA's 0.70 A/W requirements.

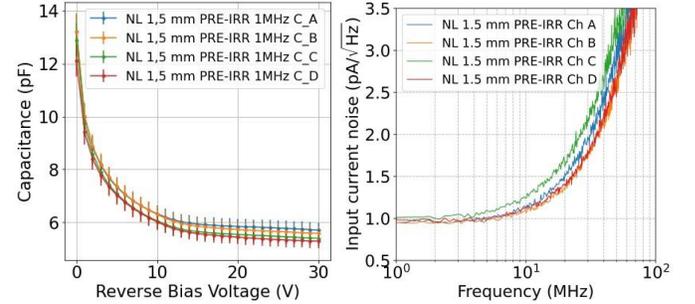

*Figure 4 Pre-irradiation capacitance of 1.5mm NL QPD (left) and input current noise for the associated QPR (right) for all channels.*

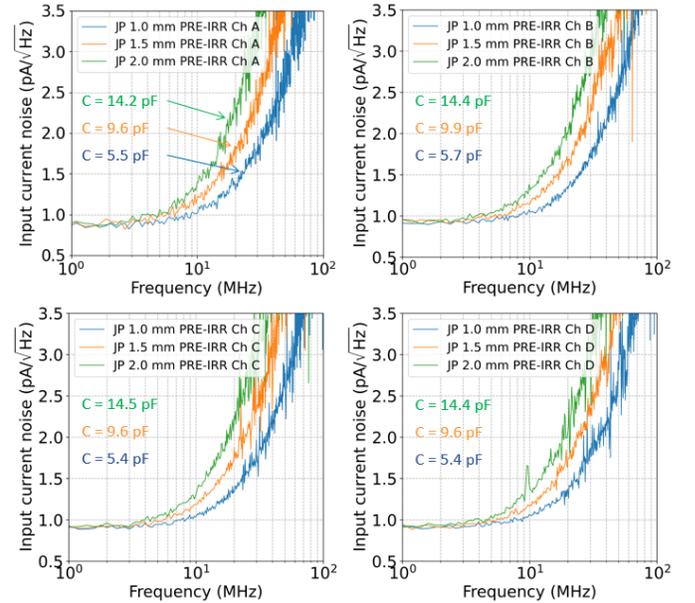

*Figure 5 Input current noise spectra of the QPR with the JP QPDs for each channel before irradiation.*

### B. Irradiation results

After all irradiations, all QPDs remain functional with no breakdowns observed. While all QPDs capacitance and noise measurements remain constant pre- and post-irradiation, only QPD with a 2.0 mm active area, irradiated with 20 MeV proton, displays a noise increase up to 0.16 pA/√Hz and a 2.0 MHz bandwidth reduction, as seen in Figure 6. This effect can be attributed to the QPR input current noise being influenced by the shot noise, which in turn is governed by both the photocurrent $i_{ph}$ and the dark current $i_{dark}$, as articulated in equation (2) [17].

$$i_{shot}(f) = 2e[I_{ph} + I_{dark}] \ (A^2/Hz) \quad (2)$$

Where $i_{shot}(f)$ represents the spectral shot noise density at the frequency f.

As elaborated later, the dark current accentuates with diameter and decreases with energy. In most QPDs, the photocurrent contribution in the noise dominates over the dark current. However, for the 2.0 mm QPD irradiated with 20 MeV protons, the significant increase in dark current results in it becoming the dominant source of noise, over the contribution from the photocurrent. This could explain the lack of noise increase in other QPDs, aligning with our theoretical model.

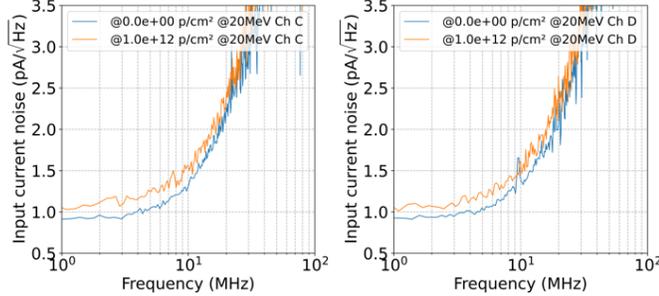

*Figure 6 Input current noise for segments C and D pre- and post-irradiation of the 2.0 mm QPD irradiated @20 MeV.*

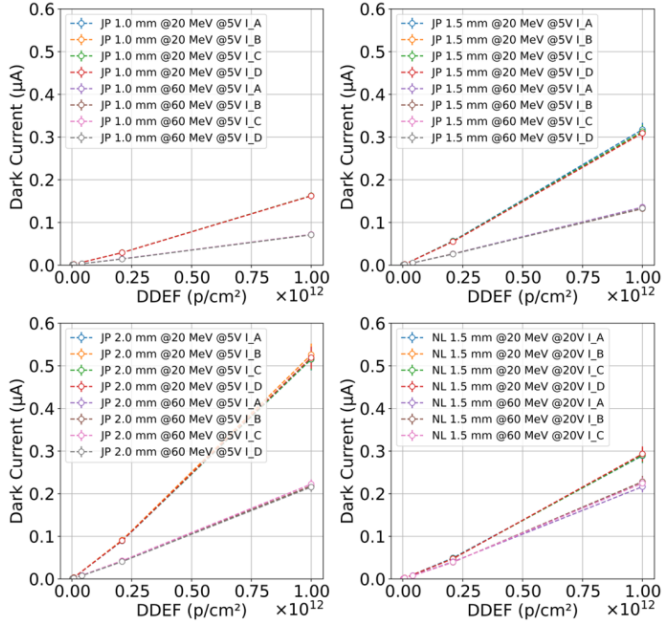

*Figure 7 Dark current progression for each quadrant (A, B, C, and D) of 1.0 mm and 2.0mm QPDs during proton irradiation @20°C.*

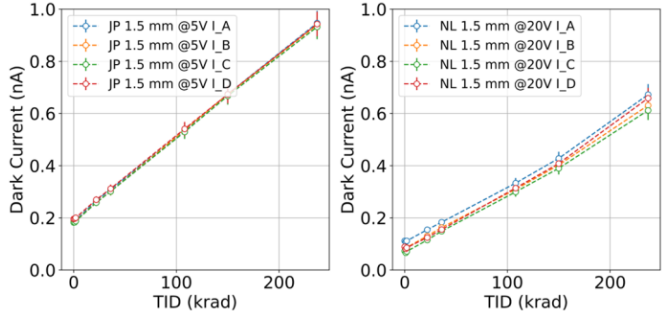

*Figure 8 Dark current progression for each quadrant (A, B, C, and D) of 1.0 mm and 2.0mm QPDs during gamma irradiation @20°C.*

The dark current consistently displays a linear growth with fluence across all irradiations, as illustrated in Figure 7 and Figure 8. The most pronounced increase was observed with 20 MeV protons, registering a maximum of 0.5 µA at 20°C and reaching up to 4.4 µA at 50°C. Notably, at the LISA fluence of $1.80 \times 10^{+11}$ p/cm², the observed levels remained beneath the LISA threshold of 1 µA for temperatures <50°C and all QPDs.

The QPD's responsivity shows a decline only after proton irradiation, but even at a fluence level of $1 \times 10^{+12}$ p/cm², it comfortably meets the LISA benchmark of 0.7 A/W. As depicted in Figure 9, the drop in responsivity post-proton irradiation is more pronounced for 20 MeV by a factor of ~1.7 compared to 60 MeV.

The drop in responsivity results in a corresponding reduction in the QPR's amplitude response to signals from an interferometric setup like LISA. This is due to the QPD's diminished photocurrent generation. However, the signal's phase remains unaltered.

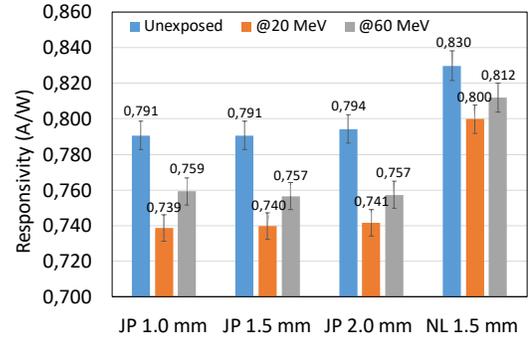

*Figure 9 Responsivity of non-irradiated QPD and irradiated QPD with 20 & 60 MeV. Note: Each point represents the average responsivity of the 4 quadrants of the QPD Error bars reflect the measurement precision of 1%.*

## IV. DAMAGE FACTOR AND DISCUSSION

To compare the performance of these detectors, the experimental damage factor $K_{Idark}$ is computed and can be seen in Figure 10. It quantifies the increase in dark current relative to fluence and is defined in equation (3).

$$K_{Idark} = \frac{I_{dark}(\Phi) - I_{dark}(0)}{\Phi \cdot S} \quad (3)$$

Where $I_{dark}(\Phi)$ is the dark current measured at the fluence $\Phi$ and $I_{dark}(0)$ the dark current measured prior irradiation. In the $K_{Idark}$ formula, the dark current is divided by the QPD sensitive aera S to get the dark current density.

It is important to highlight that the growth in dark current originates from both ionizing and non-ionizing damage mechanisms. Hence, a direct comparison with the NIEL introduces an error. Gamma radiation, predominantly characterized by ionizing damage, acts as a benchmark for discerning the ionizing damage component within proton irradiation. At a maximum fluence of $1 \times 10^{+12}$ p/cm², a total ionizing dose (TID) of 237 krad was deposited, mirroring the TID achieved with gamma irradiation. The observed increase in dark current was notably minor, being less than 1 nA (Figure 8). Therefore, ionizing damage in proton irradiation appears to be minimal (<1%) and can be overlooked.

In Figure 10, damage factors were ascertained from QPDs subjected to protons irradiation at both 20 MeV and 60 MeV.

These factors were determined after each irradiation interval. As illustrated in Figure 10, the damage factor for 60 MeV irradiation remains relatively stable, exhibiting an average increase of approximately 19% between the initial and final steps. In contrast, QPDs irradiated at 20 MeV demonstrate an escalation of approximately 51%. Ideally, at such fluence levels, damage factors should be invariant to the applied fluence. Statistical effects can lead to an increase in the damage factor with the fluence [18], [19]. However, the investigation focuses on large-diameter photodiodes, statistical fluctuations are typically discernible at lower fluence levels, well beneath $1\times10^{+12}$ p/cm² [18]. Therefore, the increment observed in this research is not attributable to such mechanisms.

As expected, the damage factors resulting from 20 MeV protons irradiation exceed that of 60 MeV protons. For JP QPDs, the dark current degradation ($K_{Idark}$) from 20 MeV is approximately double compared to 60 MeV. In contrast, for NL photodiodes the ratio is closer to 1.34, aligning more consistently with anticipated values as provided by InGaAs NIEL (NIEL ratio between 20 and 60 MeV =1.37). These discrepancies can be caused by the technology of the QPDs, but more information is needed to confirm it. Future work will focus on understanding these differences.

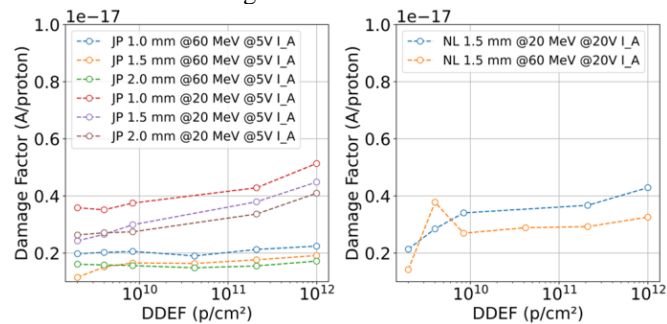

*Figure 10 Dark current damage factor calculated for different applied fluences. The left panel presents JP diodes of different diameters irradiated with 20 MeV and 60 MeV protons. The right panel presents analogous measurements for 1.5 mm NL QPD.*

## V. CONCLUSION

The InGaAs QPDs have showcased robust radiation tolerance, with no breakdown under various radiation types such as electron, gamma, and proton. Their resilience was tested up to a TID of 237 krad and a DDD of $4.9\times10^{+9}$ MeV/g. Dark current exhibited a consistent rise, peaking at ~0.5 µA for the 2 mm QPD under 20 MeV proton irradiation but remained within LISA's specifications. While capacitance and noise largely remained unchanged, a 0.16 pA/√Hz uptick was seen in the same 2 mm QPD post-20 MeV irradiation, likely due to the dark current's increment. Even with a maximum responsivity drop of 6.6%, the levels still exceed LISA's 0.7 A/W requirement, reaffirming QPDs' suitability for space applications like LISA.

There are deviations between the theoretical predictions of the NIEL and the experimental damage factor results such as the observed rise in the damage factor with fluence during 20 MeV proton irradiation. Such discrepancies, possibly linked to the QPD technology, merit more detailed exploration.

## VI. ACKNOWLEDGEMENT

The authors would like to express their gratitude to the team at CAL for their technical assistance during the proton irradiation and to Romain Rey for his technical support during electron irradiation at ONERA. The authors also extend their thanks to the ARTEMIS laboratory from the Observatories de la Côte d'Azur. In addition, the authors would like to acknowledge the LISA QPR working group for their support and insightful discussions. Their contributions have been invaluable in the development of this work.